# Abnormal enhancement of electric field inside a thin permittivity-near-zero object in free space

Yi Jin[1], Pu Zhang[1,2] and Sailing He[1,2*]

[1]*Centre for Optical and Electromagnetic Research, JORCEP [Joint Research Center of Photonics of the Royal Institute of Technology (Sweden) and Zhejiang University], State Key Laboratory of Modern Optical Instrumentations, Zhejiang University, Hangzhou 310058, China*

[2]*Department of Electromagnetic Engineering, School of Electrical Engineering, Royal Institute of Technology, 100 44 Stockholm, Sweden*

It is found that the electric field can be enhanced strongly inside a permittivity-near-zero object in free space, when the transverse cross section of the object is small and the length along the propagation direction of the incident wave is large enough as compared with the wavelength. The physical mechanism is explained in details. The incident electromagnetic energy can only flow almost normally through the outer surface into or out of the permittivity-near-zero object, which leads to large energy stream density and then strong electric field inside the object. Meanwhile, the magnetic field inside the permittivity-near-zero object may be smaller than that of the incident wave, which is also helpful for enhancing the electric field. Two permittivity-near-zero objects of simple shapes, namely, a thin cylindrical shell and a long thin rectangular bar, are chosen for numerical illustration. The enhancement of the electric field becomes stronger when the permittivity-near-zero object becomes thinner. The physical mechanism of the field enhancement is completely different from the plasmonic resonance enhancement at a metal surface.

Metamaterials can be utilized to realize many special electromagnetic responses that do no exist in nature. The current intense interest in metamaterials mainly originates from the theoretical work of Pendry *et al*. in which they provided a way to realize negative index materials (NIMs) [1,2], and the experimental work of Smith *et al*. in which they demonstrated the existence of NIMs by using arrays of metallic split ring resonators and wires at microwave frequencies [3]. Among various special material properties provided by metamaterials (i.e., negative, very small or very large permittivity and permeability), near zero permittivity or permeability is also an interesting research topic which has attracted much attention recently [4-12]. Some noble metals and polar materials can also provide near zero permittivity. Engheta *et al*. pointed out that the electromagnetic wave in a thick metallic waveguide can tunnel through a very narrow metallic waveguide [with a perfect electric conductor (PEC) boundary] filled with a permittivity-near-zero material [6]. Strong electric field can be obtained in the narrow waveguide. Later, we suggested that a dielectric split ring in a permeability-near-zero material can force all the electromagnetic energy from a central excitation source to flow through the narrow gap where large energy stream density and strong magnetic field can be obtained [12]. Here, we will show that due to some particular properties of a permittivity-near-zero material, strong electric field can appear inside a simple thin permittivity-near-zero object in free space without using a PEC boundary or any other boundary condition. A thin cylindrical shell and a long thin rectangular bar will be investigated to illustrate this interesting phenomenon. In the investigation, two-dimensional electromagnetic propagation is assumed with the magnetic field perpendicular to the *x-y* plane (TM polarization), and the time harmonic factor is *exp*(−*iωt*).

We first investigate the cylindrical shell shown in Fig. 1. Its permittivity is $\varepsilon_2$, and its inner and outer radii are $r_1$ and $r_2$, respectively. The permittivity of the core surrounded by the shell is $\varepsilon_1$, and that of the background outside the shell is $\varepsilon_3$. When $\varepsilon_2$ approaches to zero, the shell has some particular properties. The magnetic field is constant and the electric displacement field is zero at every point inside the whole shell. According to the continuity condition of the normal displacement fields at both sides of a boundary [13], the normal displacement field in the background is zero on the outer surface ($F_2$) of the shell. Consequently, the corresponding normal electric field in the background is zero (i.e., only the tangential electric field exists) on surface $F_2$, which means the energy stream (averaged in time) can only flows in or out normally through surface $F_2$. In addition, using our previous result given in Ref. [12], it is easy to show the following fact: if the excitation source is outside the shell, the energy stream is zero at every point in the core, and no energy stream flows through any part of the inner surface ($F_1$) of the shell.

The electromagnetic distribution inside the shell is interesting. Because of the rotation symmetry of the shell in Fig. 1, the magnetic and electric fields in each domain can be expanded by cylindrical harmonics [13]. When $\varepsilon_2$ approaches to zero, after some derivation, one can obtain the following contribution of each order of the incident wave to the electromagnetic field inside the shell,

$$\begin{cases} H_{2,z,l=0}(\mathbf{r}) = \dfrac{k_3[AJ_0(k_3r_2)-J_1(k_3r_2)]}{Ak_3 - k_3^2(r_2^2-r_1^2)/2r_2 - B\varepsilon_3 k_1 r_1/\varepsilon_1 r_2} a_{3,0} \\ H_{2,z,l\neq 0}(\mathbf{r}) = 0 \end{cases} \quad (1a)$$

$$\begin{cases} E_{2,r,l=0}(\mathbf{r}) = 0 \\ E_{2,r,l\neq 0}(\mathbf{r}) = -\dfrac{C_m k_3}{\omega \varepsilon_3}(-\dfrac{r}{r_2})^{m-1}\dfrac{1+(r_1/r)^{2m}}{1-(r_1/r_2)^{2m}} e^{il\theta} a_{3,l} \end{cases} \quad (1b)$$

$$\begin{cases} E_{2,\theta,l=0}(\mathbf{r}) = i\{\dfrac{\omega\mu_0 r}{2} + \dfrac{r_1}{\omega r}[\dfrac{Bk_1}{\varepsilon_1} - \dfrac{\omega^2\mu_0 r_1}{2}]\} H_{2,z,l=0}(\mathbf{r}) \\ E_{2,\theta,l\neq 0}(\mathbf{r}) = \dfrac{iC_m k_3}{\omega \varepsilon_3}(-\dfrac{r}{r_2})^{m-1}\dfrac{1-(r_1/r)^{2m}}{1-(r_1/r_2)^{2m}} e^{il\theta} a_{3,l} \end{cases} \quad (1c)$$

where $J_l(x)$ and $H_l(x)$ are the Bessel and Hankel functions of the first kind [13], $A=H_1(k_3r_2)/H_0(k_3r_2)$, $B=J_1(k_1r_1)/J_0(k_1r_1)$, $C_m=-2i/[\pi k_3 r_2 H_m(k_3 r_2)]$, $m=|l|$, $a_{3,l}$ is the expansion coefficient when the incident wave outside the shell is expanded by Bessel functions, $k_n$ is the wave number in the $n$-th domain ($n=1,2,3$) and $\mu_0$ is the vacuum permeability. One can see that $H_{2,z,l=0}(\mathbf{r})$ is independent of position $\mathbf{r}$, which is consistent with the constant magnetic field inside the shell. It is also noticed that $E_{2,r,l\neq 0}(\mathbf{r})$ and $E_{2,\theta,l}(\mathbf{r})$ are nonzero, which means that unlike the magnetic field, the electric field is inhomogeneous and its phase may differ at different positions. Most importantly, as the thickness ($d=r_2-r_1$) of the shell is very small compared to its radius, i.e., $r_1/r_2$ is about one, $E_{2,r,l\neq 0}(\mathbf{r})$ can be very large with finite $a_{3,l\neq 0}$. When $r_1/r_2 \to 1$, $\varepsilon_1=\varepsilon_3$, and the incident wave is a plane wave, one can obtain the following limiting distribution of the total electromagnetic field inside the shell,

$$H_{2,z}(\mathbf{r}) \to J_0(k_3 r_2), \quad (2a)$$

$$E_{2,r}(\mathbf{r}) \to -\dfrac{4i}{\pi\omega\varepsilon_3 d}\sum_{m=1}^{\infty}\dfrac{\cos[m(\theta-\theta_0-\pi/2)]}{mH_m(k_3 r_2)}, \quad (2b)$$

$$E_{2,\theta}(\mathbf{r}) \rightarrow \frac{ik_3}{\omega\varepsilon_3} J_1(k_3 r_2) - \frac{4}{\pi\omega\varepsilon_3 r_2}(1-\frac{r_2-r}{d})\sum_{m=1}^{\infty}\frac{\cos[m(\theta-\theta_0-\pi/2)]}{H_m(k_3 r_2)}, \quad (2c)$$

where $\theta_0$ is the angle between the propagation direction of the plane wave and the $+x$ axis. As shown in Eq. (2b), as $r_1/r_2 \rightarrow 1$, the total longitudinal electric field, $E_{2,r}(\mathbf{r})$, is independent of $r$, and will become infinitely large, which is inversely proportional to $d$. And as shown in Eq. (2c), the total transverse electric field, $E_{2,\theta}(\mathbf{r})$, is dependent on $r$, which means that it is always inhomogeneous along any cross section of the shell as $r_1/r_2 \rightarrow 1$. Unlike $E_{2,r}(\mathbf{r})$, $E_{2,\theta}(\mathbf{r})$ is finite as $r_1/r_2 \rightarrow 1$. Thus, inside a very thin shell, the energy stream propagates mainly along the shell since the magnetic field is perpendicular to the *x-y* plane. The above particular phenomenon does not exist in a structure formed with conventional materials.

As a numerical example, we choose a plane wave propagating along the $+x$ axis. Both the core and background are assumed to be air. When $r_1=\lambda_0$ ($\lambda_0$ is the wavelength in vacuum) and shell thickness $d=0.15\lambda_0$, Fig. 2(a) shows the distributions of the electric field amplitude and the energy stream density around the shell. In all out numerical results, an electromagnetic quantity is always normalized by the absolute value of the corresponding quantity of the incident plane wave. As shown in Fig. 2(a), when the plane wave impinges on the shell, the energy stream flows normally through surface $F_2$ into the shell, but it can not continue to flow through surface $F_1$ into the core and has to flow along the shell. The energy stream inside the shell gradually accumulates to a peaky value as polar angle $\theta$ decreases from 180° toward 0°, and thereafter, it gradually leaks into the outer background. Since the magnetic field is constant inside the shell, the distribution character of the electric field amplitude is nearly the same as that of the absolute energy stream density, but not completely the same as the phase of the electric field differs at different positions.

Now we study the influence of the thickness of the shell. As $d$ decreases, the electric field inside the shell is gradually enhanced and can be arbitrarily large, as expected from Eq. (1b). Fig. 3(a) shows electric field amplitude $|\mathbf{E}_p|$ at point $p(r=r_1+d/2, \theta=\theta_m)$ for different $d$ by curve 1. $p$ is such a point at which the electric field amplitude is maximal on the middle semi-circle between surfaces $F_1$ and $F_2$ ($0°\leq\theta\leq180°$). One can see clearly that $|\mathbf{E}_p|$ is inversely proportional to $d$ approximately. This can be explained as follows. For a thin shell, if $d$ is reduced further, the electromagnetic field outside the shell will change little as it can hardly sense such a reduction. Thus, the energy stream flowing normally through the outer surface of the thin shell is kept nearly the same, and so is the total energy stream through any cross section of the shell. As shown by curve 5 in Fig. 3(a), the total energy stream through the cross section connecting point $p_1(r=r_1, \theta=\theta_m)$ and point $p_2(r=r_2, \theta=\theta_m)$ decreases very little as $d$ decreases. Then, the energy stream density inside the shell has to increase. Absolute energy stream density $|\mathbf{S}_p|$ at point $p$ is shown by curve 2 in Fig. 3(a), which is inversely proportional to $d$ approximately. Although only a small part of the incident energy enters the shell, most of which is reflected by the shell, very large energy stream density can still be obtained inside the shell if the shell is enough thin. In the mean time, as $d$ decreases, the constant magnetic field inside the shell becomes only a bit weaker gradually as shown by curve 3 in Fig. 3(a) (weaker than that of the incident wave). Such variation tendencies of the energy stream density and magnetic field lead to the fast enhancement of the electric field inside the shell as its thickness decreases. The normalized value of curve 1 is larger than that of curve 2 for any $d$ in Fig. 3(a), and the ratio of curve 2 to curve 1 is shown by the curve in the inset of Fig. 3(a), which is similar to curve 3. It should be noted that for $\theta$ near 0° or near 180°, the electric field can not be very large even if the shell is very thin as shown by curves 1 and 3 in Fig. 2(b). This is because when $\theta$ is near 180°, the energy stream density inside the shell has not accumulated enough to a large value, and when $\theta$ is near 0°, most of the energy has leaked away from the shell as shown by curves 2 and 4 in Fig. 2(b). Since the magnetic field is constant inside the shell, small energy stream density means small electric field at a position.

We then study the influence of the radius of the shell. Inside a shell of larger radius, although the magnetic field is constant, the electric field and energy stream density exhibit their oscillating character more obviously. In Fig. 2(b), more oscillating peaks appear on curves 3 and 4 as compared with curves 1 and 2. Fig. 3(b) shows $|\mathbf{E}_p|$, $|\mathbf{S}_p|$ and $|\mathbf{H}_p|$ at point $p$ as $r_1$ increases. At some special values of $r_1$, $|\mathbf{H}_p|$ and $|\mathbf{S}_p|$ tend to zero, whereas $|\mathbf{E}_p|$ has no such tendency. This can be understood in the following way. Magnetic field $\mathbf{H}_1(\mathbf{r})$ in the core is determined by $a_{1,0}J_0(k_1r)$, where $a_{1,0}$ is always about equal to $a_{3,0}$ when enlarging the thin shell. $J_0(k_1r)$ is an oscillating function possessing zero points. When $k_1r_1$ approaches to a zero root of $J_0(k_1r)$, $J_0(k_1r_1)$ tends to zero. Thus, when enlarging the shell, $\mathbf{H}_1(\mathbf{r})$ may tend to zero near surface $F_1$ at some special values of $r_1$. According to the continuity of the tangential magnetic fields at both sides of a boundary [13], constant magnetic field $\mathbf{H}_2(\mathbf{r})$ inside the shell also tends to zero. This leads to a scenario that no incident energy enters the shell where the energy stream density is zero. However, the electric field inside the shell may not tend to zero, since the zero magnetic field can assure the zero energy stream density. In addition, when $k_1 r_1$ approaches to infinity, $J_0(k_1 r_1)$ tends to zero and then the magnetic field inside the shell also gradually tends to zero as shown by the decreasing peaks of curve 3 in Fig. 3(b). This makes the energy stream through a local area on surface $F_2$ smaller, but the incident energy can enter the shell on a larger total area, and the maximal accumulated energy stream density inside the shell can still grow slowly as shown by the slowly increasing peaks of curve 2 in Fig. 3(b). The variation tendencies of the magnetic field and energy stream density lead to the gradual enhancement of the electric field (curve 1 of Fig. 3(b)) inside the shell when $r_1$ increases, as expected from Eq. (1b).

In the above simulation, the shell is assumed to be made of a permittivity-zero material. In fact, a permittivity-near-zero material would have some material loss in general. Fig. 4 shows $|\mathbf{E}_p|$ at point $p$ for different $d$ and $r_1$ when some material loss is introduced. Material loss can degenerate the strong enhancement effect of the electric field inside the shell. As shown in Fig. 4(a), when the loss is large (i.e., $\varepsilon_2/\varepsilon_0 = 10^{-2} i$), the electric field inside the shell can not become infinitely large as the thickness of the shell approaches zero. And as shown in Fig. 4(b), when the loss is large, the electric field inside the shell can not be strongly enhanced by enlarging the dimension of the shell with the thickness fixed. However, if the loss is moderate (i.e., $\varepsilon_2/\varepsilon_0=10^{-4}i$), strong electric field can still be obtained inside the shell. In practice, to make the absolute imaginary part of $\varepsilon_2$ small, low-loss metamaterials need to be intelligently designed, or introduction of gain is an effective method.

Similar strong enhancement of the electric field can also be obtained inside a long thin rectangular permittivity-near-zero bar shown in Fig. 5(a). When the bar in air is of width $d=0.05\lambda_0$, length $l=\lambda_0$, permittivity $\varepsilon_1/\varepsilon_0=10^{-4}i$, and a plane wave propagates along the $+x$ axis, Fig. 5(b) shows the distributions of electric field amplitude and energy stream density calculated with a finite-element-method software, COMSOL [14]. Fig. 5(c) shows $|\mathbf{E}_p|$, $|\mathbf{S}_p|$ and $|\mathbf{H}_p|$ at point $p$ as thickness $d$ decreases. $p$ is such a point at which the electric field amplitude is maximal on the middle axis parallel to the $x$ axis. Similar to the shell in Fig. 1, when the bar is enough thin, the electromagnetic field outside the bar is nearly independent of its thickness. Thus, the energy stream flowing through the surface of the bar changes little as the bar is thinned, and then the energy stream density inside the bar is inversely proportional to $d$ approximately as shown by curve 2 in Fig. 5(c). Meanwhile, similar to the shell case, the magnetic field inside the bar is also weaker than that of the incident wave as shown by curve 3 in Fig. 5(c), and becomes a bit weaker slowly as thickness $d$ is reduced. The above tendencies of the energy stream density and magnetic field lead to the fast enhancement of the electric field inside the bar when thickness *d* decreases, as shown by curve 1 in Fig. 5(c). To obtain a strong electric field, the bar needs to be long to make more incident energy enter the bar. However, when the bar is long enough, the electric field inside the bar can not be further enhanced as shown by curve 1 in Fig. 5(d). If the plane wave propagates along the +*y* axis instead of the +*x* axis, the scattering problem can be approximately treated as a problem of reflection and transmission of a thin film inside which strong electric field enhancement can not be obtained.

In conclusion, we have shown that strong electric field can be obtained inside a thin cylindrical shell or a long thin rectangular bar of near zero permittivity. In the present field enhancement mechanism, no plasmonic resonance is used (and thus the maximal field occurs in the middle, instead of near the surface in a typical surface plasmon resonance), and the key physical property is that the energy stream can only flow nearly normally through the outer surface into or out of the permittivity-near-zero object. It is also helpful for the enhancement of the electric field that the magnetic field in the permittivity-near-zero object is weaker than that of the incident wave. The enhancement of the electric field can also appear inside other permittivity-near-zero shapes when the cross section transverse to the propagation direction of the incident wave is small and the length along the propagation direction of the incident wave is large enough as compared with the wavelength. Unlike earlier studies on a permittivity-near-zero waveguides by Engheta *et al*. [6-10], the present approach does not require a PEC boundary or any other boundary condition, and the permittivity-near-zero object is in free space. Strong electric field has many interesting applications in e.g. sensing and some nonlinear problems. For convenience of utilizing the strong electric field, a permittivity-near-zero material can be constructed with a multilayered structure consisting of negative permittivity layers (i.e., noble metals) and positive permittivity layers (i.e., air or other common dielectrics). The sensing or nonlinear processes can occur in the positive-permittivity layers. With such a method, the operating wavelength can be flexibly tuned. With a similar method, strong magnetic field enhancement can be achieved inside an open thin permeability-near-zero object, as one would expect from the duality of Maxwell's equations.

This work is partially supported by the National Natural Science Foundation (Nos. 60990322 and 60901039) of China, and the Swedish Research Council (VR) and AOARD.

**Figure captions :**

FIG. 1. Configuration of a cylindrical shell.

FIG. 2. (Color Online) (a) Distributions of the electric field amplitude and energy stream density (shown in arrows) when a plane wave propagating along the +$x$ axis impinges on a permittivity-zero shell with $r_1=\lambda_0$ and $d=0.15\lambda_0$. (b) Electric field amplitude $|\mathbf{E}_s(\theta)|$ (curves 1 and 3) and absolute energy stream density $|\mathbf{S}_s(\theta)|$ (curves 2 and 4) on the middle semi-circle between surfaces $F_1$ and $F_2$ when $d=0.01\lambda_0$. Curves 1 and 2 are for $r_1=\lambda_0$, and curves 3 and 4 are for $r_1=3\lambda_0$.

FIG. 3. (Color Online) Variations of some electromagnetic quantities inside a permittivity-near-zero shell when (a) $d$ decreases with fixed $r_1=\lambda_0$, and (b) $r_1$ increases with fixed $d=0.01\lambda_0$. Curves 1−3 are for electric amplitude $|\mathbf{E}_p|$, absolute energy stream density $|\mathbf{S}_p|$ and magnetic amplitude $|\mathbf{H}_p|$ at point $p(r=r_1+d/2, \theta=\theta_m)$, respectively. In (a), curve 4 is for absolute total energy stream $|\mathbf{P}_p|$ through the cross section connecting points $p_1(r=r_1, \theta=\theta_m)$ and point $p_2(r=r_2, \theta=\theta_m)$, and the ratio of curve 2 to curve 1 is shown in the inset.

FIG. 4. (Color Online) Variation of $|\mathbf{E}_p|$ inside a permittivity-near-zero shell when material loss is introduced. (a) $d$ decreases with fixed $r_1=\lambda_0$, and (b) $r_1$ increases with fixed $d=0.01\lambda_0$. Curves 1−4 are for $\varepsilon_2/\varepsilon_0=0$, $10^{-4}i$, $10^{-3}i$, and $10^{-2}i$, respectively. The sampled positions for $|\mathbf{E}_p|$ when loss is introduced are the same as that when there is no loss.

FIG. 5. (Color Online) (a) Configuration of a rectangular bar. (b) Distributions of the electric field amplitude and energy stream density (shown in arrows) when a plane wave propagating along the $+x$ axis impinges on the bar with $\varepsilon_1/\varepsilon_0=10^{-4}i$, $d=0.05\lambda_0$ and $l=\lambda_0$. (c) Variations of some electromagnetic quantities as thickness $d$ increases with fixed length $l=\lambda_0$, and (d) those as length $l$ increases with fixed thickness $d=0.01\lambda_0$. In (c) and (d), curves 1−3 are for $|\mathbf{E}_p|$, $|\mathbf{S}_p|$ and $|\mathbf{H}_p|$ at point $p$, respectively.

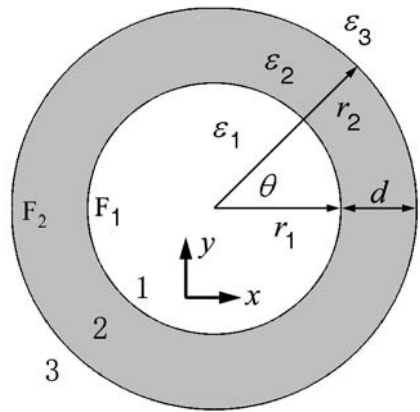

FIG. 1. Configuration of a cylindrical shell.

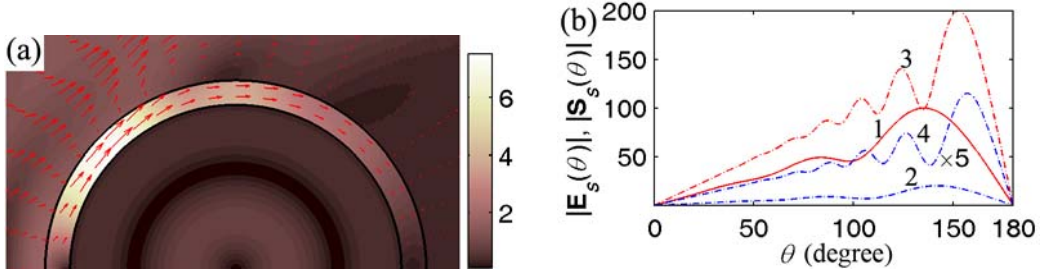

FIG. 2. (Color Online) (a) Distributions of the electric field amplitude and energy stream density (shown in arrows) when a plane wave propagating along the $+x$ axis impinges on a permittivity-zero shell with $r_1=\lambda_0$ and $d=0.15\lambda_0$. (b) Electric field amplitude $|\mathbf{E}_s(\theta)|$ (curves 1 and 3) and absolute energy stream density $|\mathbf{S}_s(\theta)|$ (curves 2 and 4) on the middle semi-circle between surfaces $F_1$ and $F_2$ when $d=0.01\lambda_0$. Curves 1 and 2 are for $r_1=\lambda_0$, and curves 3 and 4 are for $r_1=3\lambda_0$.

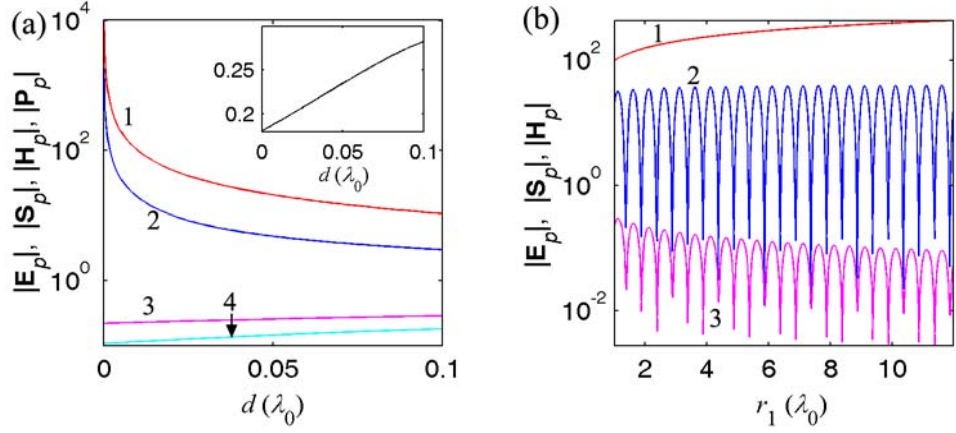

FIG. 3. (Color Online) Variations of some electromagnetic quantities inside a permittivity-near-zero shell when (a) $d$ decreases with fixed $r_1=\lambda_0$, and (b) $r_1$ increases with fixed $d=0.01\lambda_0$. Curves 1−3 are for electric amplitude $|\mathbf{E}_p|$, absolute energy stream density $|\mathbf{S}_p|$ and magnetic amplitude $|\mathbf{H}_p|$ at point $p(r=r_1+d/2, \theta=\theta_m)$, respectively. In (a), curve 4 is for absolute total energy stream $|\mathbf{P}_p|$ through the cross section connecting points $p_1(r=r_1, \theta=\theta_m)$ and point $p_2(r=r_2, \theta=\theta_m)$, and the ratio of curve 2 to curve 1 is shown in the inset.

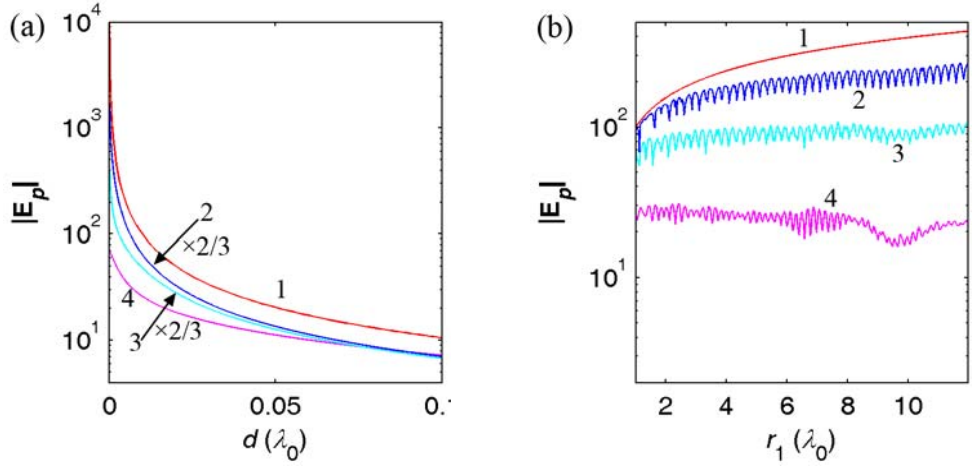

FIG. 4. (Color Online) Variation of $|\mathbf{E}_p|$ inside a permittivity-near-zero shell when material loss is introduced. (a) $d$ decreases with fixed $r_1=\lambda_0$, and (b) $r_1$ increases with fixed $d=0.01\lambda_0$. Curves 1−4 are for $\varepsilon_2/\varepsilon_0=0$, $10^{-4}i$, $10^{-3}i$, and $10^{-2}i$, respectively. The sampled positions for $|\mathbf{E}_p|$ when loss is introduced are the same as that when there is no loss.

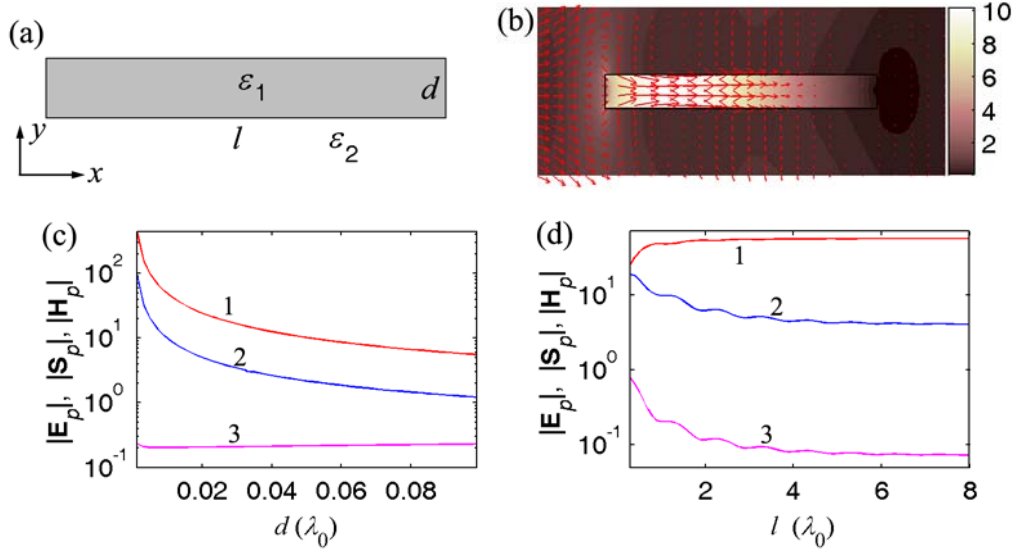

FIG. 5. (Color Online) (a) Configuration of a rectangular bar. (b) Distributions of the electric field amplitude and energy stream density (shown in arrows) when a plane wave propagating along the $+x$ axis impinges on the bar with $\varepsilon_1/\varepsilon_0=10^{-4}i$, $d=0.05\lambda_0$ and $l=\lambda_0$. (c) Variations of some electromagnetic quantities as thickness $d$ increases with fixed length $l=\lambda_0$, and (d) those as length $l$ increases with fixed thickness $d=0.01\lambda_0$. In (c) and (d), curves 1−3 are for $|\mathbf{E}_p|$, $|\mathbf{S}_p|$ and $|\mathbf{H}_p|$ at point $p$, respectively.